\documentclass[
  journal=pasa,
  manuscript=research-paper, 
  year=2022,
  volume=XX,
]{cup-journal}

\usepackage{microtype,siunitx,booktabs,amssymb,amsmath}
\title{A Search for Annihilating Dark Matter in 47 Tucanae and Omega Centauri}

\author{Lister Staveley-Smith}
\affiliation{International Centre for Radio Astronomy Research (ICRAR), University of Western Australia, 35 Stirling Highway, Crawley, WA 6009, Australia}
\alsoaffiliation{ARC Centre of Excellence for All Sky Astrophysics (ASTRO 3D), Australia}
\email[Lister Staveley-Smith]{Lister.Staveley-Smith@uwa.edu.au}

\author{Emma Bond}
\affiliation{International Centre for Radio Astronomy Research (ICRAR), University of Western Australia, 35 Stirling Highway, Crawley, WA 6009, Australia}

\author{Kenji Bekki}
\affiliation{International Centre for Radio Astronomy Research (ICRAR), University of Western Australia, 35 Stirling Highway, Crawley, WA 6009, Australia}

\author{Tobias Westmeier}
\affiliation{International Centre for Radio Astronomy Research (ICRAR), University of Western Australia, 35 Stirling Highway, Crawley, WA 6009, Australia}
\alsoaffiliation{ARC Centre of Excellence for All Sky Astrophysics (ASTRO 3D), Australia}

\received {dd Mmm YYYY}
\revised  {dd Mmm YYYY}
\accepted {dd Mmm YYYY}
\published{XX Month 202X}


\begin{document}
\setlength\emergencystretch\columnwidth


\begin{abstract}
A plausible formation scenario for the Galactic globular clusters 47 Tucanae (47 Tuc) and Omega Centauri ($\omega$ Cen) is that they are tidally stripped remnants of dwarf galaxies, in which case they are likely to have retained a fraction of their dark matter cores. In this study, we have used the ultra-wide band receiver on the Parkes telescope (Murriyang) to place upper limits on the annihilation rate  of exotic Light Dark Matter particles ($\chi$) via the $\chi\chi\rightarrow e^+e^-$ channel using measurements of the recombination rate of positronium (Ps).
This is an extension of a technique previously used  to search for Ps in the Galactic Centre. However, by stacking of spectral data at multiple line frequencies, we have been able to improve sensitivity.
Our measurements have resulted in 3-$\sigma$ flux density (recombination rate) upper limits of 1.7 mJy ($1.4\times 10^{43}$~s$^{-1}$) and 0.8 mJy  ($1.1 \times 10^{43}$ s$^{-1}$) for 47 Tuc and $\omega$ Cen, respectively. 
Within the Parkes beam at the cluster distances, which varies from 10--23 pc depending on the frequency of the recombination line,  and for an assumed annihilation cross section $\langle\sigma v\rangle = 3\times 10^{-29}$ cm$^3$~s$^{-1}$, we calculate upper limits to the dark matter mass and rms dark matter density of  $\lesssim 1.2-1.3\times 10^5 f_n^{-0.5}$ ($m_\chi/$MeV~c$^{-2}$) M$_{\odot}$ and $\lesssim 48-54 f_n^{-0.5}$ ($m_\chi/$MeV~c$^{-2}$) M$_{\odot}$ pc$^{-3}$ for the clusters, where $f_n=R_n/R_p$ is the ratio of Ps recombination transitions to annihilations, estimated to be $\sim 0.01$.
The radio limits for $\omega$ Cen suggest that, for a fiducial dark/luminous mass ratio of $\sim0.05$, any contribution from Light Dark Matter is small unless $\langle\sigma v\rangle <  7.9\times 10^{-28}$ ($m_\chi/$MeV~c$^{-2}$)$^2$ cm$^3$ s$^{-1}$. Owing to the compactness and proximity of the clusters, archival 511-keV measurements suggest even tighter limits than permitted by CMB anisotropies, $\langle\sigma v\rangle < 8.6\times 10^{-31}$ ($m_\chi/$MeV~c$^{-2}$)$^2$ cm$^3$ s$^{-1}$.
Due to the very low synchrotron radiation background, our recombination rate limits substantially improve on previous  radio limits for the Milky Way.
\end{abstract}

\section{INTRODUCTION}

The nature of dark matter remains mysterious, even though studies of galaxies and the Cosmic Microwave background radiation suggest it makes up 26\% of the cosmic mass-energy density and 84\% of the density of normal matter in the present-day Universe \citep{Planck2018}. However, precise knowledge of its phase-space distribution, its self-interaction,  and its coupling to normal matter is important for understanding the formation and subsequent evolution of galaxies. 

Whilst macroscopic scenarios such as Primordial Black Holes \citep{Niikura2019} or massive compact halo objects \citep{Tisserand2007} remain possible dark matter candidates, the available parameter space is small, and most modern experiments have focused on attempted detection of Weakly Interacting Massive Particles (WIMPs), which are cold, dissipationless non-standard particles \citep{Jungman1996}. However, attempts to detect WIMPs have also so far failed, resulting in renewed exploration of other dark matter particle candidates, particularly Light Dark Matter (LDM)  \citep{Boehm2004b,Knapen2017} and axions \citep{PQ1977,ADMX2021}, though many other dark matter candidates have been suggested, including sterile neutrinos \citep{Adhikari2017}, Kaluza-Klein particles \citep{Servant2003}, Strongly Interacting Massive Particles (SIMPs) \citep{Hochberg2015}, and Fuzzy Dark Matter \citep{Hui2017}.  For comprehensive reviews of the dark matter problem, particle candidates and detection schemes, see \citet{Bergstrom1998rev}, \citet{Garrett2011} and \citet{Bertone2018}.

Astronomical searches for the electromagnetic signature of dark matter through their decay or annihilation products (particularly WIMPs, LDM and axions) have been conducted at many wavelengths from $\gamma$-rays to the radio regime, and over size scales from the magnetospheres of individual pulsars (a few 10's of km) to the largest virialised structures in the Universe -- clusters of galaxies. 

Searches at the high energies (0.1-100 TeV) with the High Energy Stereoscopic System (H.E.S.S.) have been used to constrain the cross section of self-annihilating WIMPS in the Galactic Centre and nearby dwarf galaxies \citep{Abdallah2018,Abdallah2021}. Such searches are sensitive to mono-energetic photon annihilation products arising from the $\chi\chi\rightarrow \gamma\gamma$ annihilation channel as well as to the $\gamma$-ray continuum resulting from a wide range of other ($W^+W^-$, $Z^+Z^-$, $b\bar{b}$, $t\bar{t}$, $e^+e^-$, $\mu^+\mu^-$ and $\tau^+\tau^-$)  annihilation channels available in Supersymmetric Standard Models \citep[see][]{Bergstrom1998th}. At lower energies (0.1--300 GeV), Fermi-LAT $\gamma$-ray observations of the Galactic Centre, galaxies and clusters of galaxies have similarly been used to constrain the annihilation cross section of lower-mass WIMPs \citep{Bringmann2012,Ackermann2015,Ackermann2017,Thorpe-Morgan2021}. For the Galactic Centre, annihilation cross section upper limits in the range $\langle\sigma v\rangle<10^{-26}$ to $10^{-25}$ cm$^3$ s$^{-1}$ in the mass range 10~GeV $< m_{\chi}c^2 < 50$ TeV are typical. The Cherenkov Telescope Array should improve on these limits in certain mass ranges \citep{Silverwood2015} .

Radio observations are also sensitive to WIMP GeV annihilation products due to the synchrotron radiation emitted by the acceleration of charged particles (e.g. protons, electrons and their anti-particles) in pre-existing magnetic fields \citep{Borriello2009}. Observations of nearby galaxies have resulted in similarly useful annihilation cross section limits of around $\langle\sigma v\rangle<10^{-26}$ to $10^{-25}$ cm$^3$ s$^{-1}$ in the mass range 1-1000 GeV c$^{-2}$ \citep{Regis2017,Cook2020,Basu2021}, though with a dependence on other parameters such as magnetic field strength, diffusion coefficients and equipartition \citep{Colafrancesco2006}.
Future radio observations, including with the Square Kilometre Array, are expected to substantially improve on these limits \citep{Storm2017,ChenSKA2021}.

Self-annihilating LDM particles in the 1-1000 MeV~c$^{-2}$ mass range can produce lower-energy $\gamma$-rays, including from the $e^+e^-$ annihilation line at 511 keV. Although this line is strongly detected in the Galactic Centre \citep{Johnson1973,Leventhal1978,SiegertMW2016}, its origin is unclear as other $e^+$ production channels exist, including pair production around pulsars \citep{Jones1979} and $\beta^+$ decay in supernovae \citep{Crocker2017,SiegertMW2022}. Observations of nearby galaxies have not yet revealed any significant 511 keV detections, but have nevertheless resulted in excellent  cross-section limits, especially for low mass LDM particles. For example, for $m_{\chi}\sim 1$ MeV~c$^{-2}$, \citet{SiegertRet2022} deduce a limit of $\langle\sigma v\rangle< 5\times 10^{-28}$ cm$^3$ s$^{-1}$ for the dwarf galaxy Reticulum II (Ret II).

Axions, if they exist, are likely to have a mass in the range $1-1000 \mu$eV c$^{-2}$ which puts their energy range firmly in the radio regime. Via the Primakoff effect \citep{Kelley2017,Millar2021}, axions can convert into a photon in a strong magnetic field. Recent observations, including those of the magnetar PSR J1745-2900 by \citet{Darling2020}, have found no narrow spectral features which could be attributed to axion conversion. Laboratory measurements \citep{ADMX2021} and further radio observations with sensitive telescopes at radio-quiet sites \citep{Wang2021} are likely to make major improvements in sensitivity and mass range in the near future.

The most likely places to observe electromagnetic signatures of dark matter are in dense dark matter cores, which are predicted to occur in the nuclei of galaxies, and should have size scales in the range 0.1 to 5 kpc \citep{Lazar2020}. Thus searches have concentrated on objects such as the Galactic Centre \citep{SiegertMW2016,Reynolds_2017}, the Large Magellanic Cloud \citep[LMC;][]{Siffert2011} and nearby dwarf galaxies \citep{SiegertSat2016,Albert2017,Cook2020,Abdallah2021,SiegertRet2022}. The dwarf galaxies have the advantage of being more dark-matter dominated, with fewer sources of strong emission which arise from particle acceleration from other astrophysical processes such as star formation, supernovae and massive black holes. For example, the strong detection of the 511 keV annihilation line in the Milky Way may have a nucleosynthetic origin, rather than a dark matter origin \citep{SiegertMW2022}. However, a disadvantage is that their distances (typically 20 to 500 kpc) are mostly larger than for the Milky Way or LMC, leading to weaker signal strength predictions.

Another avenue are the much closer giant globular clusters 47 Tucanae (47 Tuc) and Omega Centauri ($\omega$ Cen), also known as NGC 5139. These objects have distances of 4.4 kpc and 5.5 kpc, respectively \citep{Chen_2018,Del_Principe_2006} and may be the nucleated remnants of ancient tidally-stripped dwarf galaxies \citep{1984ApJ...277..470P,Norris1996,hilker_richtler_2000,Bekki_2006,Lee_2009,2021MNRAS.500.4578M} and therefore in possible possession of a dark matter core. { Various authors have suggested the role that dark matter may play in explaining their velocity dispersion profiles \citep[e.g.][]{Penarrubia2017}. \citet{Evans2022} have used GAIA, HST and spectroscopic data to infer strong evidence for a non-luminous component in $\omega$ Cen, with a  mass contained within the half-light radius of up to $10^6$~M$_{\odot}$. No 511 keV radiation has been detected from these clusters \citep{Knodlseder2005}, though observations with Fermi-LAT have detected strong $\gamma$-ray continuum \citep{Abdo2010}.} Due to the poor angular resolution of $\gamma$-ray telescopes, the source of the emission is not clear -- it is consistent with models of the { annihilation} of massive dark matter particles \citep{Gaskins2016,brown2019glow,Wirth2020}, but could also arise from a population of old millisecond pulsars which is known to exist in these clusters \citep{Abdo2010,Reynoso_Cordova_2019}. There are 25 known millisecond pulsars in 47 Tuc \citep{Freire_2017} and 5 in $\omega$ Cen \citep{Dai_2020}, plus a number of unidentified X-ray sources with similar properties to millisecond pulsars \citep{Bhattacharya2017,Henleywillis2018}.

 \citet{Wirth2020} used high-resolution computer simulations to model the evolution of nucleated dwarf galaxies near the Milky Way and found that they can transform into globular clusters similar to $\omega$ Cen. Their simulated nuclei were found to have central dark matter densities of 0.1 to several M$_{\odot}$ pc$^{-3}$.  \citet{Wirth2020} were able to match the predicted $\gamma$-ray spectrum (assuming dark matter { annihilation} via the $b\bar{b}$ channel) of one of their simulated clusters with that observed by Fermi-LAT for $\omega$ Cen using an annihilation cross section   $\langle\sigma v\rangle = 5 \times 10^{-28}$ cm$^3$ s$^{-1}$.

Whether or not the $\gamma$-ray detections of 47 Tuc and $\omega$ Cen represent the signature of dark matter { annihilation}, they at least give an upper limit to the dark matter { annihilation} rate, under the assumption that their cores really do contain dark matter. This applies to scenarios of WIMP { annihilation}, as explored by \citet{Wirth2020}, as well as { annihilation} of LDM and other candidate dark matter particles. 

In this paper, we introduce a search for signatures of $e^+e^-$ pair production resulting from the  annihilation of LDM particles in these globular clusters by using observations of positronium (Ps) radio recombination lines. Radio observations have the advantage of potentially much higher angular resolution than $\gamma$-ray observations, and therefore the potential to discriminate between different astrophysical origins. Furthermore, spectral-line observations also have the potential to reduce confusion with diffuse and/or compact sources of emission in the foreground and background. 

The existence of copious amounts of Ps atoms is undisputed in the centre and plane of the Milky Way. { As already mentioned, the $\gamma$-ray annihilation line at 511 keV is strongly detected and has been known for many decades, with detailed inspection of the continuum \citep{SiegertMW2016} suggesting that the line mostly arises from Ps annihilation, rather than direct $e^+e^-$ annihilation.
Corresponding radio searches for Ps recombination lines towards the} Galactic Centre have been conducted by \citet{Ananth1989} and \citet{Reynolds_2017}.
The search by \citet{Reynolds_2017} was more sensitive to the diffuse emission expected from dark matter annihilation and stellar populations. They established a 3-$\sigma$ upper limit to the brightness temperature of any Ps recombination line emission of $T_B < 0.09$ K. Due to the strong synchrotron continuum background { and factors discussed later,} this limit is still well above the rate measured by $\gamma$-ray observations of $2 \times 10^{43}$ s$^{-1}$ \citep{SiegertMW2016}.

In this paper, Section~\ref{sec:positronium} summarises the expected characteristics of Ps recombination lines in the astrophysical context; Section~\ref{sec:observations} describes the details of the new globular cluster observations and the instrument details; Section~\ref{sec:analysis} describes the data analysis, calibration, stacking methodology and results; Section~\ref{sec:discussion} discusses the results and their implications for the LDM self-annihilating dark matter content of the clusters;  and Section~\ref{sec:summary} provides a summary and a few comments about future research directions.

\section{POSITRONIUM (Ps) RECOMBINATION}
\label{sec:positronium}

Ps atoms, which consist of an electron ($e^-$) and a positron ($e^+$), can be formed from an $e^-/e^+/$hydrogen (H) plasma once energy losses in the interstellar medium have reduced the gas temperature to $T_k < 10^6$ K \citep{Burdyuzha1994}. At these temperatures, Ps can form from recombination of electrons and positrons, or by charge exchange of positrons with H and other atoms and molecules, with the relative formation rates being extremely sensitive to temperature and density \citep{Bussard1979}. Below $10^4$ K, radiative recombination becomes dominant, leading to the greater possibility of the formation of Ps atoms in excited states \citep{Wallyn1996}. The key ingredient for this paper is that once Ps atoms are formed in an excited electronic state, energy loss will proceed by a cascade of emission of hydrogenic-like recombination lines, followed by rapid annihilation near the ground state.

Ps annihilation is very rapid, with a timescale of $1.25\times10^{-10} n^3$ s for para-Ps and $1.33\times 10^{-7} n^3$ s for ortho-Ps \citep{Burdyuzha1996}, where $n$ is the principal quantum number. Selection rules only allow for annhilation in angular momentum $\ell=0$ states, which will depopulate low-$n$ energy levels, especially for the shorter-life para-Ps species, but will not depopulate higher levels, each of which also has $n$ angular momentum states $\ell=0,1,...,n-1$.

As laboratory measurements have shown \citep{Canter1975}, the Ps recombination spectrum is analogous to the hydrogen spectrum in Galactic HII regions, which is dominated by Ly$\alpha$ at ultraviolet wavelengths, H$\alpha$ and H$\beta$ in the optical and high-order recombination lines at radio wavelengths. Ps line intensities are similar to those of hydrogen \citep{Wallyn1996}.

The general Rydberg formula for the rest frequency of recombination lines from hydrogenic-like atoms is:

\begin{equation}
\nu = \mathrm{c}RZ^2\left(\frac{1}{n_1^2}-\frac{1}{n_2^2}\right) ,
\label{eq:rydberg}
\end{equation}
where c is the velocity of light, $R$ is the Rydberg constant, $Z$ is the atomic charge, $n_1$ is the principal quantum number of the lower energy level, and $n_2$ is the principal quantum number of the upper energy level.

The nomenclature for Ps recombination lines follows standard practice for electronic radio-frequency transitions \citep{Lilly1968}, where a line is labelled by the chemical element followed by the principal quantum number of the lower energy level of the transition. The change in level $\Delta n=n_2-n_1=1,2,$... is denoted by Greek letters $\alpha$, $\beta$,..., so that Ps93$\alpha$ represents a transition from principle electronic quantum number $n_2=94$ to $n_1=93$. Missing Greek letters are usually taken to imply $\Delta n =1$. 

The Rydberg constant for Ps is $R_{Ps} = m_e e^4/(16\epsilon_0^2 h^3 c) = 5.48686578 \times 10^6$ m$^{-1}$, where $m_e$ and $e$ are the electronic mass and charge respectively, $\epsilon_0$ is the permittivity of free space and $h$ is the Planck constant. Thus $R_{Ps}$ is approximately half the value of $R_H$, corresponding to the ratio of the respective reduced masses of Ps and H. Therefore Ps recombination lines for a given level will have approximately half the frequency (and half the transition probability) of the corresponding H lines. As an example, the rest frequency of Ps93$\alpha$ is 4024.99 MHz.

 Line shapes and widths in the astrophysical context depend on numerous parameters such as temperature, velocity dispersion, pressure, transition probability, and optical depth. For hydrogen radio recombination lines in the warm, diffuse interstellar medium, Gaussian broadening by a combination of temperature and velocity dispersion are usually dominant. However, being a light atom, Ps lines are expected to be dominated by thermal broadening. The usual formula for line-of-sight non-relativistic thermal broadening in frequency is:
 \begin{equation}
\Delta \nu = \nu \sqrt{\frac{2\mathrm{k}T}{m \mathrm{c}^2}} ,
\label{eq:thermal}
\end{equation}
where $\nu$ is the line frequency, $k$ is the Bolzmann constant, $T$ is temperature and $m$ is the mass of the chemical species. The thermal linewidths of Ps lines are therefore expected to be $\sqrt{m_H/2m_e} \approx 30.3$ times broader than those of hydrogen lines.  For an electron temperature $T_e = 5\times 10^3$ K, the expected Ps frequency dispersion is $\Delta \nu = 3.69$ MHz for a Ps93$\alpha$ line at rest, or a FWHP linewidth of 8.69 MHz. The corresponding thermal velocity dispersion is $\sigma_v = 275$ km s$^{-1}$.

The recombination line brightness spectrum is normally a function of optical depth, radiation field, density, volume, temperature (including the effect of any spatial variations within the beam) and thermodynamic equilibrium \citep{Peters2012}. For radio recombination lines (large $n$), the spontaneous emission rate $A_{n+1,n}$, is less than the typical collision frequency $C_n$ in the diffuse ISM. For example, $A_{94,93}({\rm Ps93}\alpha)\approx 0.4$ s$^{-1}$, and the corresponding $\Delta n=1$ cascade time is $\Sigma_{n=1}^{93} A^{-1}_{n+1,n} \approx 40$ s. This compares to Ps collision frequency of $\sim 10^{-8} n^4$ s$^{-1}$ at a temperature of $5\times 10^3$ K and density of 1 cm$^{-3}$ (with substantially lower densities expected for an LDM $e^+e^-$ plasma). So line emission cannot be maintained by low-$n$ Ps collisions in the low-density ISM  as is the case for the $\lambda$21-cm line.

Line emission could arise from stimulated emission and absorption. In fact, this is the normal situation for radio recombination lines \citep{Gordon2002}. However, such emission relies on the region being illuminated by continuum radiation. This can arise from synchrotron emission or thermal free-free emission at the frequencies considered here. But, given the low densities expected and the fact there is no evidence for strong diffuse radio continuum emission in 47 Tuc and $\omega$ Cen, this is unlikely to be the case here.

As previously noted \citep{Ananth1989,Ananth1993,Reynolds_2017}, the main emission source is therefore likely to be the radiative recombination cascade that immediately follows the formation of a Ps atom.
The intensity spectrum will reflect the distribution of energy levels following Ps formation, which depends sensitively on temperature. \cite{Burdyuzha1996} cite $P(n)\propto n^{\alpha}$ with $\alpha=-1$ at low temperature and $\alpha=-3$ at extremely high temperatures, with the detailed calculation of \citet{Wallyn1996} suggesting $\alpha\approx -2$ at 5000 K (very different from the Saha collisional index $\alpha=2$). 
Recombination will then proceed via the more probable $\Delta n=1$ transitions, so that every large $n$ state will give rise to $\Delta n=1$ transitions at most lower energy levels. A reasonable approximation, and an upper limit to the $\Delta n=1$ photon emission spectrum is therefore $I_{\gamma}(n)\propto n^{\alpha+1}$, rather than the $n^0$ spectrum assumed previously \citep{Ananth1989,Ananth1993,Reynolds_2017}.\footnote{By considering all possible recombination and annihilation cascades and extrapolating from $n<15$, Table~6 of  \citep{Wallyn1996} suggests a pessimistic lower limit to the low-temperature high-$n$ emission spectrum of $I_{\gamma}(n)\propto n^{\alpha-1}$.}
For $T=5000$ K and $\alpha=-2$, the $\Delta n=1$ photon spectrum will be approximated by $I_{\gamma}(n)\propto n^{-1}$, corresponding to a flux spectrum $I_{\nu}(n)\propto \nu^{4/3}$ (from Equation~\ref{eq:rydberg}) and a peak flux density spectrum $S_{\nu}\propto \nu^{1/3}$ (from Equation~\ref{eq:thermal}).

In addition to uncertainties in $\alpha$, there are a number of other factors which may affect Ps recombination line intensities:

\begin{enumerate}
    \item Direct annihilation: detailed inspection of the Ps $\gamma$-ray annihilation spectrum of the Galactic Centre suggests that virtually all $e^+$ annihilation is via the Ps atomic state, rather than direct annihilation \citep{SiegertMW2016}.  Nevertheless, at temperatures above 20,000 K, the latter is important \citep{Wallyn1996}.
    \item Charge exchange: if H or He atoms are present, formation of Ps by charge exchange \citep{Guessoum2005} will reduce the recombination line intensities, as the Ps atoms will be preferentially created in low-$n$ states \citep{Wallyn1996}. This is again likely to be an important factor only at higher temperatures ($>5000$ K), and is highly dependent on the $e^+/\textrm{H}$ abundance.
    \item Dust: \citet{Guessoum2010} point out that $e^+$ annihilation on small dust grains can non-negligibly reduce the Ps fraction in the warm ionised interstellar medium.
    \item Higher-order lines: recombination lines with $\Delta n>1$ weaken as $\Delta n^2$ \citep{Gordon2002}, but provide a non-negligible contribution to radiative energy loss. Fine and hyperfine transitions are similarly weak \citep{Wallyn1996}. 
\end{enumerate}

For the purposes of later analysis, we shall characterise the ratio of the $\Delta n=1$ recombination line photon rate to the ortho+para annihilation rate by  $f_n=R_n/R_p$. As discussed above, the main contribution to $f_n$ is likely to be the Ps$n\alpha/$Ps1$\alpha$ photon ratio of $\sim n^{-1}$. However, since the value of $f_n$ is uncertain, we include it as a variable in later calculations.

\section{OBSERVATIONS}
\label{sec:observations}

We obtained data from the 64-m Parkes radio telescope (Murriyang) in NSW, Australia using the ultra-wide-bandwidth low-frequency receiver (UWL) \citep{Hobbs2020}. This receiver allowed us to survey with a higher sensitivity and over a wider instantaneous bandwidth (0.704 -- 4.032 GHz) than previously possible with this telescope. The system equivalent flux density $S_{\rm sys}$, was measured to be approximately 34 Jy at 2 GHz, but rises at both low and high frequency \citep{Hobbs2020}.

\begin{figure}
    \centering
    \includegraphics[width=1.1\textwidth]{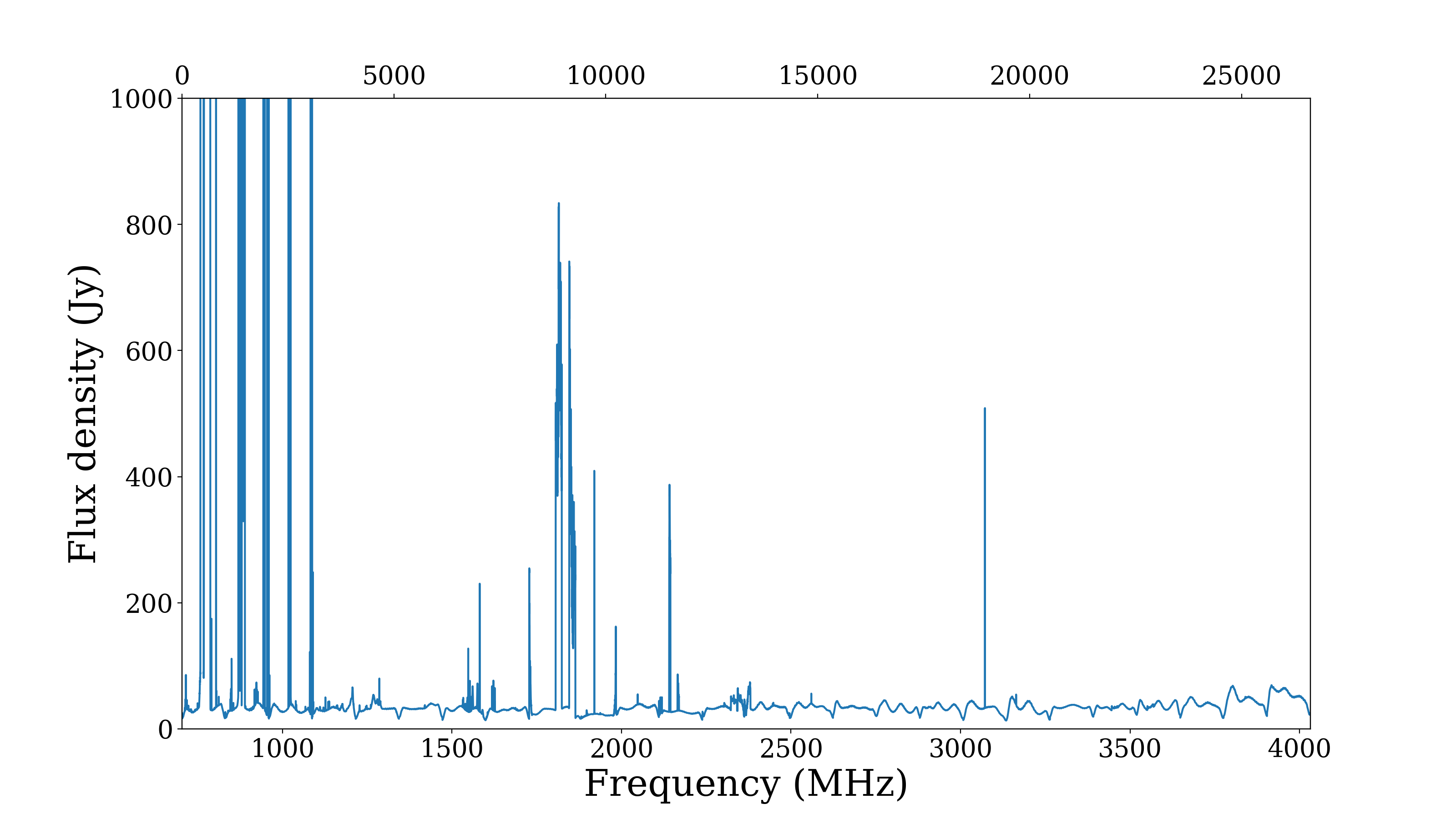}
    \caption{An example 300-s Stokes $I$ total power spectrum  taken toward 47 Tuc on 2020 October 22 at UT 13:54. The spectrum is smoothed in frequency by 32 channels to a resolution of 125 kHz and shows the locations of major RFI signals. The off-scale maximum RFI signal is $2.5\times10^6$ Jy at 762 MHz. The low-level structure in the spectrum in the regions free from RFI reflects the bandpass responses of the individual 128-MHz sub-bands. }
    \label{fig:47Tuc_TP_full}
\end{figure}

The radio-frequency interference (RFI) environment at Parkes is challenging compared to the new radio-quiet zones such as the Murchison Radio Astronomy Observatory. Figure~\ref{fig:47Tuc_TP_full} shows an example total power spectrum. About 4\% of the band contains RFI at extreme levels (more than double $S_{\rm sys}$), and around 20\% of the band contains residual RFI, after bandpass calibration, in excess of 1\% of $S_{\rm sys}$, a level which still makes the data unusable (theoretical sensitivity is exceeded in a matter of seconds). \citet{Hobbs2020} cite a similar unusable fraction (24\%) over the whole band. 

The observations were free from the 1 MHz beamformer artefacts present in previous Parkes observations \citep{Reynolds_2017}, but did suffer from filter artefacts due to the division of the band into sub-bands, each of width 128 MHz (see below, and Figures~\ref{fig:47Tuc_TP_full} and \ref{fig:waterfall}). For pulsar processing, \citet{Hobbs2020} remove a further 260 MHz (8\%) of the total bandwidth due to sub-band edge effects.

To cover the full frequency range, there were 26 sub-bands each containing 32,768 channels of width 3.9 kHz, corresponding to a velocity resolution of 0.5 km s$^{-1}$ at 2.3 GHz. Given the expected broad Ps linewidths, Doppler tracking was not enabled during the observations. 

We observed the globular clusters 47 Tuc and $\omega$ Cen over ten separate days (see Table~\ref{targets}), obtaining 24.6 hours of data for 47 Tuc and 20.3 h of data for $\omega$ Cen. The observations were mostly at night. We also observed the continuum calibrator sources listed in Table~\ref{targets}  to calibrate the flux density scale using on-off spectra. The (frequency-dependent) difference in power detected by the telescope between calibrator (on) and reference (off) positions was compared with the receiver noise power (reference position) and with an internal calibration noise diode signal. Over the elevation and time range for our daily observations, the receiver gain was remarkably stable so the noise diode was eventually powered off for most of our observations. As the noise diode and receiver noise spectra were not identical, this led to a slight improvement in bandpass calibration with little discernible change in calibration accuracy.

\begin{table}
    \centering
    \begin{tabular}{lcccc}
    \hline
    Date      & Integration & Target & Calibrator & Noise  \\
              &      h      &        &            & diode              \\
    \hline

    2020 Oct 05 & 5.1 & 47 Tuc & B1934-638 & $\checkmark$ \\
    2020 Oct 20 & 7.8 & 47 Tuc & B1934-638 & $\checkmark$ \\
    2020 Oct 21 & 3.5 & 47 Tuc & B0407-658 & $\checkmark$ \\
    2020 Oct 22 & 3.6 & 47 Tuc & B0407-658 & $\times$ \\
    2020 Oct 25 & 4.6 & 47 Tuc & B0407-658, Hydra A, & $\times$ \\
                & &        & B1934-638 & $\times$ \\
    2021 Feb 06 & 3.3 & $\omega$ Cen & Hydra A & $\times$ \\
    2021 Feb 07 & 2.6 & $\omega$ Cen & Hydra A & $\times$ \\
    2021 Feb 14 & 2.3 & $\omega$ Cen & Hydra A & $\times$ \\
    2021 Mar 20 & 6.0 & $\omega$ Cen & Hydra A & $\times$ \\
    2021 Mar 26 & 6.1 & $\omega$ Cen & Hydra A, B1934-638 & $\times$ \\

    \hline
   
    \end{tabular}
    \caption{Observations dates, integration time (on and off source), target globular cluster, calibrator and noise diode details.}
    \label{targets}
\end{table}

The globular cluster observations were made at RA\, (J2000) $00^{\rm h}24^{\rm m}05.36^{\rm s}$, Dec\, (J2000) $-72^{\circ}04'53.2''$ for 47 Tuc, and RA\, (J2000) $13^{\rm h}26^{\rm m}45.89^{\rm s}$, Dec\, (J2000) $-47^{\circ}28'36.7''$ for $\omega$ Cen.  The reference observations were made at an RA which was higher that the target RA by $5^{\rm m}$ for 47 Tuc and $3.5^{\rm m}$ for $\omega$ Cen. Since the telescope beamwidth varies from $5'$ to $26'$ over the observed frequency range \citep{Hobbs2020}, this ensured that the reference positions were more than a beamwidth away from the target for all but the very lowest frequencies (which were in any case not used in subsequent analysis due to RFI). On- and off-source observations each lasted 300 s for 47 Tuc and 180 s for $\omega$ Cen. Spectra were dumped every 1\, s into an hdf5 format data file. Both linear ($X$ and $Y$) polarisations were recorded as well as the cross-product $XY^*$.

Hydra A has a 1 GHz flux density of 58.5 Jy and a spectral index $\alpha=-0.91$ ($S_{\nu}\propto \nu^{\alpha}$) \citep{Baars1977}. B0407-658 has a 1 GHz flux density of 21.8 Jy and $\alpha=-1.28$. B1934-638 has a 1 GHz flux density of 14.8 Jy and a spectral index and curvature given by \citet{Reynolds1994}. 

\section{ANALYSIS AND RESULTS}
\label{sec:analysis}

\begin{figure}
    \centering
    \includegraphics[width=1.1\textwidth]{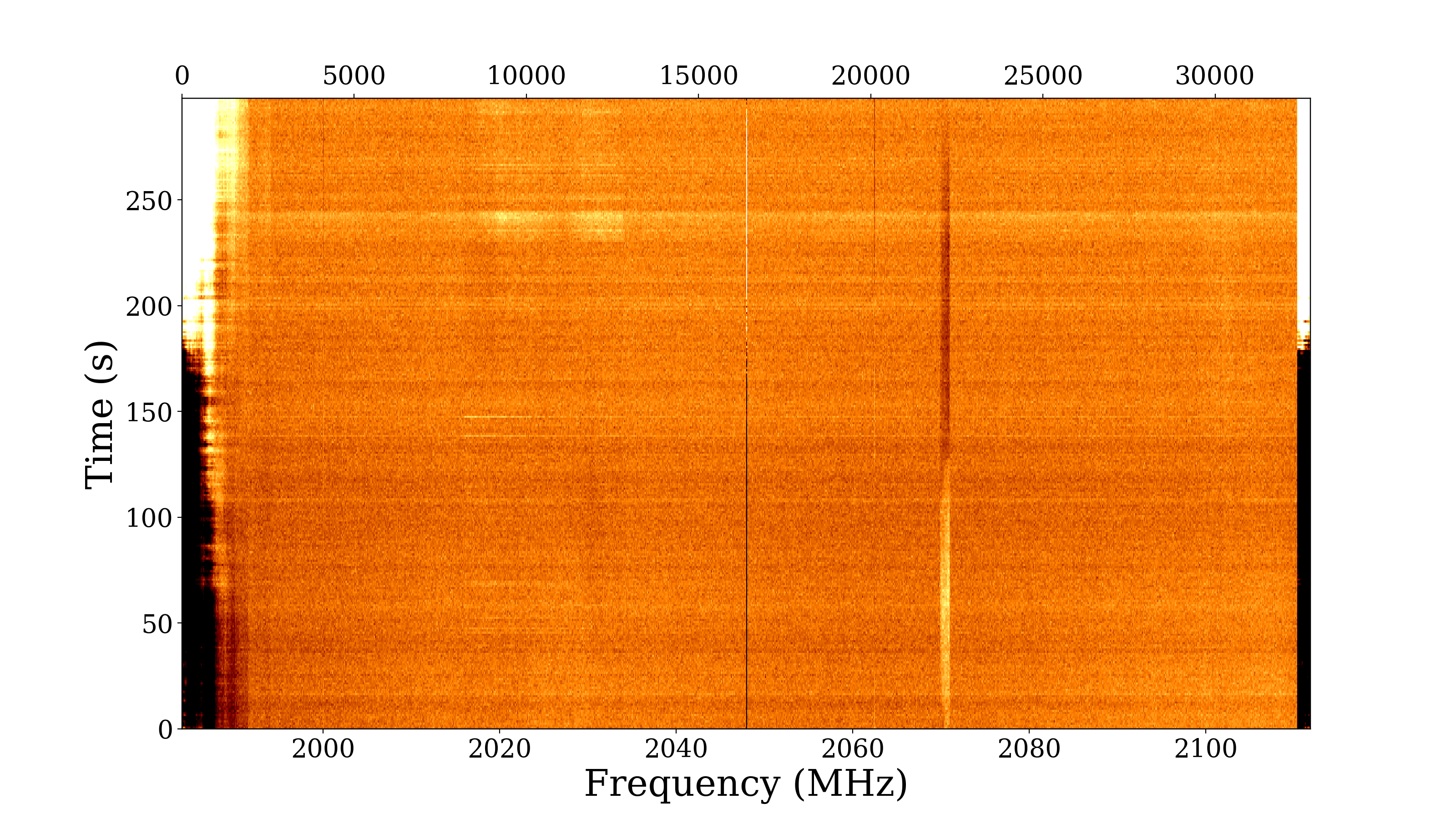}
    \includegraphics[width=1.1\textwidth]{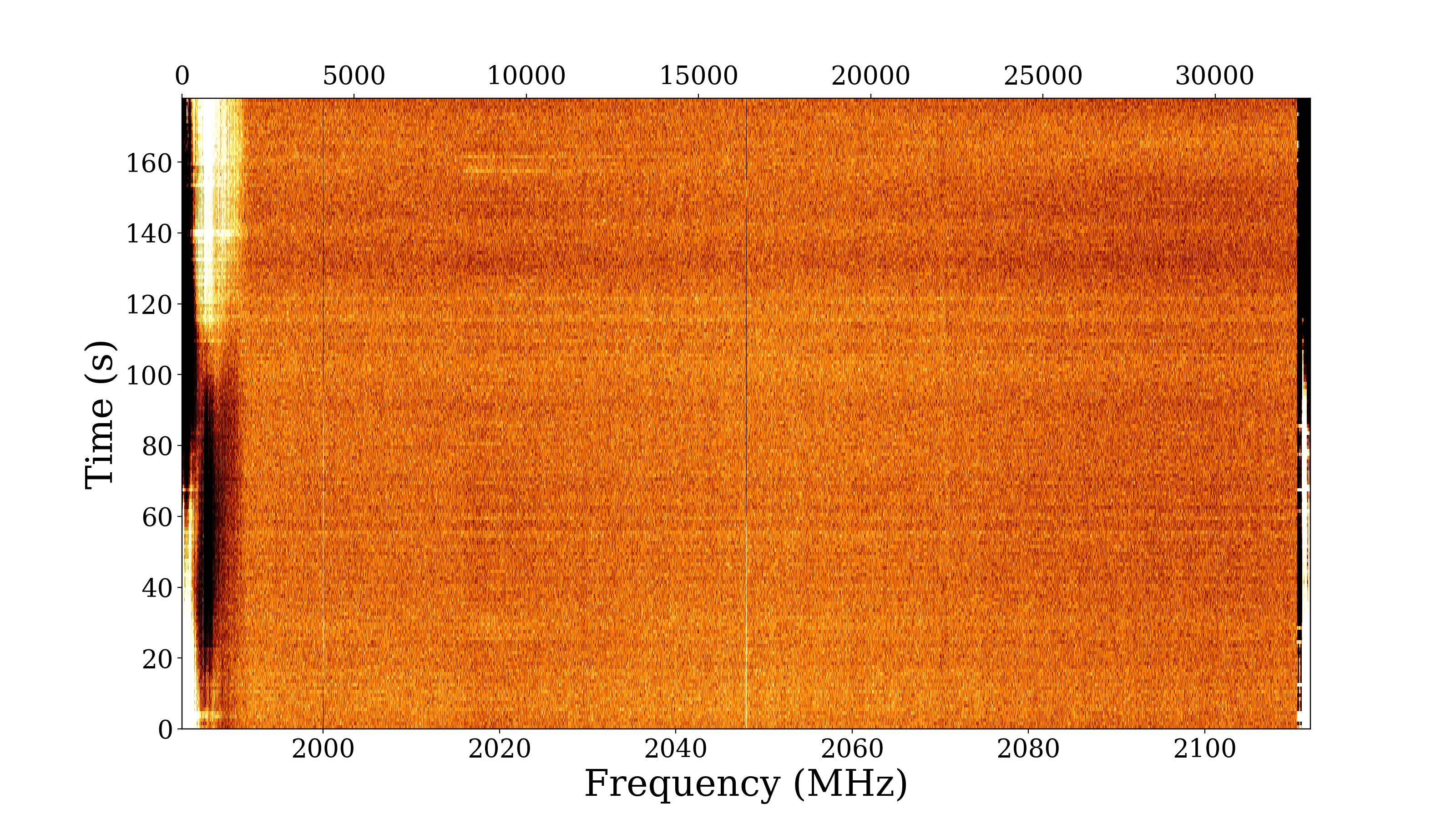}
    \caption{Example waterfall plot (time v frequency) for (top) 47 Tuc and (bottom) $\omega$ Cen where the intensity in the calibrated (quotient) Stokes $I$ spectra is displayed as a function of time and frequency for sub-band 10 for a 300-s observation taken on UT 13:54 on 2020 October 22 and UT 13:34 on 2021 February 6, respectively. Channel number is shown on the top axis. The feature at 2048 MHz is an instrumental artefact. The features at 1984 and 2112 MHz are band-edge artefacts. RFI can be seen in the ranges 2015--2034 MHz and 2070--2071 MHz. The rest-frame frequencies for three hydrogen recombination lines  (H146,147,148$\alpha$ at 2091.54,2049.29,2008.16 MHz, respectively) and two Ps recombination lines (Ps116,117$\alpha$ at 2080.72,2028.04 MHz, respectively) lie within the frequency range displayed. The intensity range is $-$1 to 1 Jy.}
    \label{fig:waterfall}
\end{figure}

The data were averaged in time across each observation (300 s for 47 Tuc and 180 s for $\omega$ Cen), and averaged in frequency by 8 channels to a resolution of 31 kHz. Bandpass calibration was achieved by forming an on-off quotient spectrum, scaled by the frequency-dependent system temperature in Jy as measured for the reference spectrum. The frequency-dependence of the system temperature was accounted for, although it was assumed to be constant across each 128 MHz sub-band. Example waterfall plots are shown in Figure~\ref{fig:waterfall}. Data away from the RFI regions were satisfactory in quality. The data for $\omega$~Cen were of slightly higher quality (flatter spectral baselines), reflecting the smaller time and angular separation of the source and reference observations. 

Sub-bands where the measured calibration was not consistent with neighbouring sub-bands were dropped  (SB8, 9 and 11). This was due to time-variable RFI affecting the calibration. The data for each cluster were then median-averaged in time. The frequency range around the predicted position for each Psn$\alpha$ line was then inspected visually. The principal quantum numbers of lines which were too close to sub-band edges, or too close to RFI were then noted, and this data was excluded from further analysis. The remainder of the data were then combined again, but this time using a weighted median across both time and quantum number. That is, all data for a given frequency offset from each Psn$\alpha$ line and at all times were median-averaged. There are 73 Psn$\alpha$ recombination lines within the frequency range of the UWL receiver. After the above exclusions, 20 Ps recombination lines were available for 47 Tuc, and 26 lines for $\omega$ Cen (see Table~\ref{stacking-list}).

\begin{table}
    \centering
    \begin{tabular}{ll}
    \hline
    Target      & Line list   \\
              &           \\
    \hline

    47 Tuc & Ps93, Ps94,  Ps96,  Ps99,  Ps100, Ps101, Ps102, Ps103, Ps105,\\
    
          & Ps106, Ps107, Ps108, Ps115, Ps116, Ps117, Ps124, Ps125, \\
            
          & Ps131, Ps132, Ps135 \\
    
    $\omega$ Cen & Ps93, Ps94, Ps97, Ps98, Ps99, Ps100, Ps101, Ps102, Ps103,\\
    & Ps104, Ps105, Ps106, Ps107, Ps108, Ps109, Ps113, Ps116, \\
    & Ps117, Ps124,Ps125, Ps129, Ps131, Ps132, Ps133, Ps134,\\
    & Ps135\\
    & \\
    \hline
   
    \end{tabular}
    \caption{A list of Psn$\alpha$ recombination lines that were able to be median-stacked for the two Globular clusters. Other line locations were not stacked due to the proximity of band edges or RFI, or poor spectral baselines. The rest-frame transition frequencies are given by Equation~\ref{eq:rydberg}. Ps93$\alpha$ lies at 4024.99 MHz; Ps135$\alpha$ lies at 1322.42 MHz.
    }
    \label{stacking-list}
\end{table}

This robust method of data combination proved to be more effective than simple averaging or clipped averaging. The method is a robust version of the spectral stacking methods employed for  $\lambda$21-cm HI and HI/HII recombination-line observations  \citep{Delhaize2013,Emig2019,Chen2021}. As with those methods, spectra were stacked in rest-frame velocity space rather than frequency space, as the predicted width of the Ps lines varies by a factor of 3 across the frequency range of the lines listed in Table~\ref{stacking-list}. This was achieved by linearly re-sampling all spectra ({\tt scipy.interpolate.interp1d}) \citep{SciPy2020}. Following the reasoning in Section~\ref{sec:positronium}, the spectra extracted around each Ps$\alpha$ line frequency were also scaled by $\nu_n^{-1/3}$ relative to the central Ps$\alpha$ line frequency (Ps113$\alpha$ or Ps114$\alpha$) to account for the increase in flux density with frequency expected for a low-density recombination cascade. Finally, to account for the variable noise levels at different times and frequencies (due to changes in RFI, receiver noise, amplitude scaling, frequency smoothing), the median absolute deviation (MAD) of the individual spectra in the central 32 MHz around each Ps$\alpha$ line frequency was calculated. The inverse square of the MAD was used as weight estimator. 

The weighted median averaging process also helped to suppress  two particular instrumental ripple artefacts: (a) a 5.7 MHz standing wave, common to the Parkes telescope \citep{Reynolds_2017}, but higher with the UWL than other receivers; and (b) a $\sim64$ MHz ripple arising from the UWL 128-MHz sub-band filters. These artefacts manifest themselves in the calibrated spectra when there are small changes in receiver gain or changes in continuum flux and spectral index between source and reference positions.

\begin{figure}
    \centering
    \includegraphics[width=1.1\textwidth]{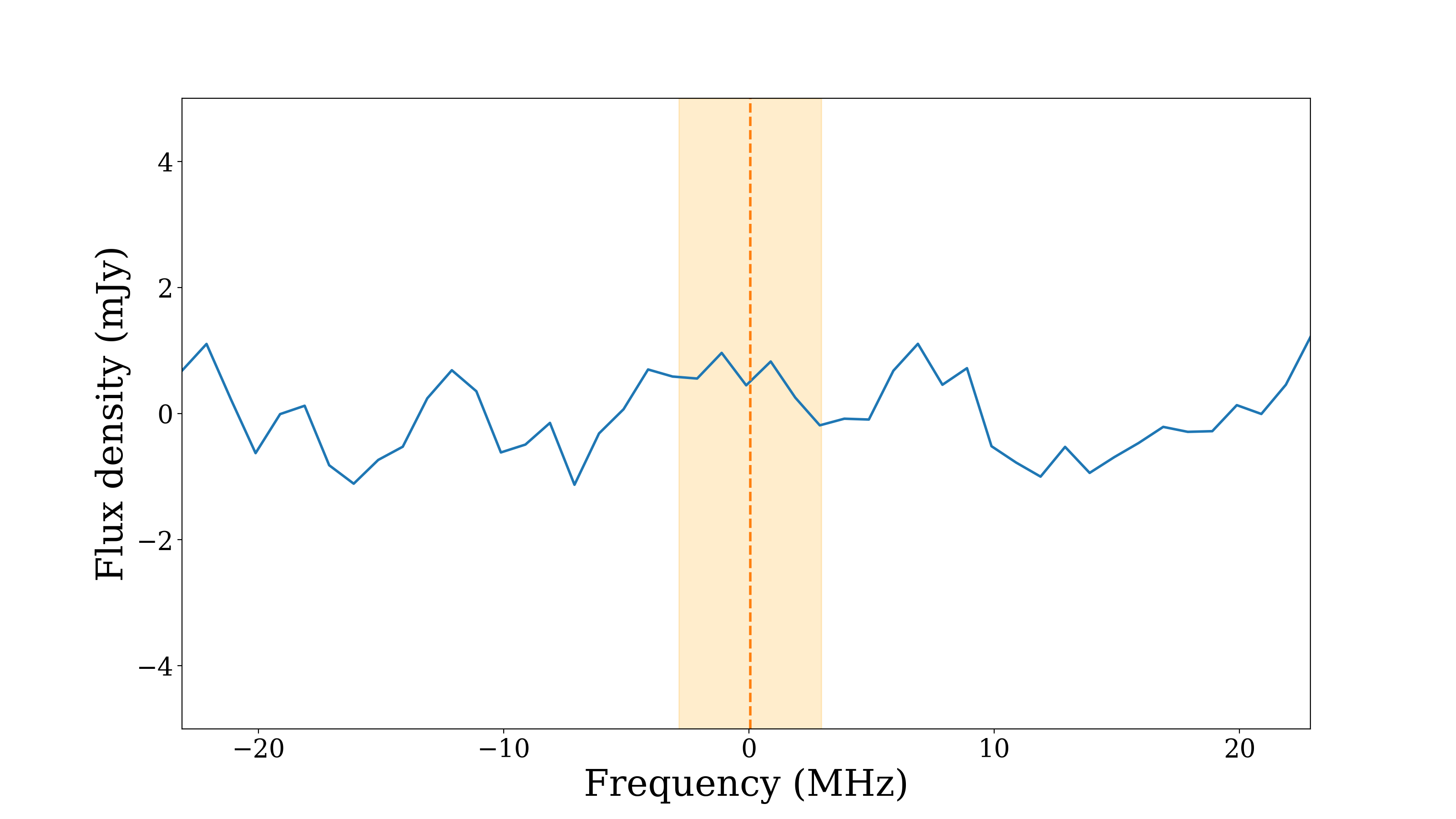}
    \includegraphics[width=1.1\textwidth]{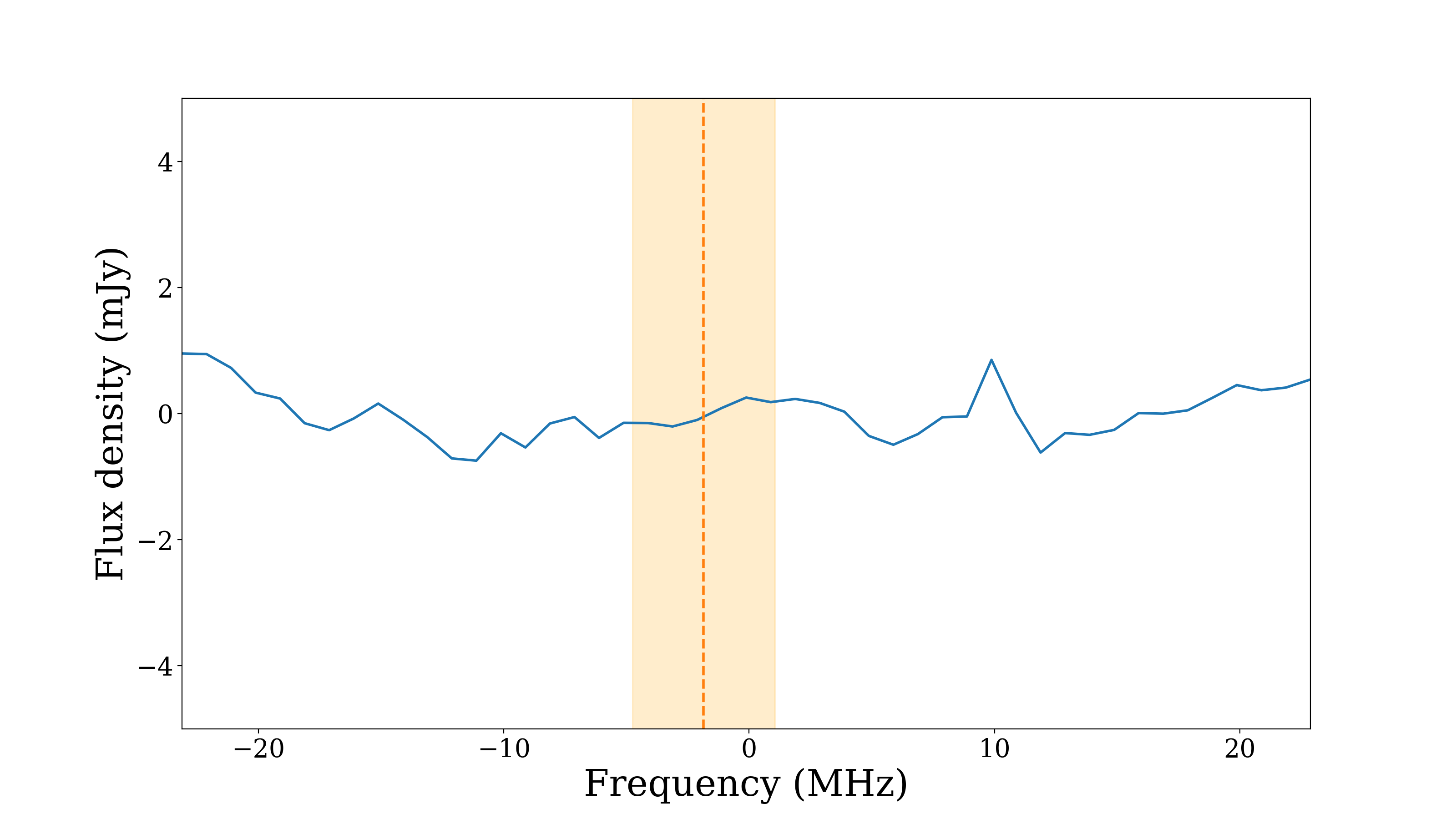}
     \caption{Stacked spectrum across all observations and all useful Psn$\alpha$ lines (see Table~\ref{stacking-list}) for (top) 47 Tuc and (bottom) $\omega$ Cen, smoothed to a frequency resolution of 1 MHz. The frequency scale is relative to zero redshift (negative frequencies correspond to redshifted emission) at  reference central frequencies of 2723.61 MHz (Ps106$\alpha$) for 47 Tuc and 2648.31 MHz (Ps107$\alpha$) for $\omega$ Cen. The expected line locations are marked with the vertical dashed lines and the expected FWHP linewidths are marked by the orange bands.}
    \label{fig:smooth8x32}
\end{figure}

The final weighted median-stacked spectra for 47 Tuc and $\omega$~Cen are shown in Figure~\ref{fig:smooth8x32}. The spectra have been smoothed to a frequency resolution of 1 MHz to more closely match the expected linewidths, and fitted with polynomials of order 3 and 1, respectively, in order to remove any residual bandpass response. The product of the number of lines stacked and the on+off source integration time is  489 and 528 h, respectively. No clear detections are made at the expected frequency offset for  the stacked spectra or for the individual recombination lines. The rms of the central 50\% of the spectra shown in Figure~\ref{fig:smooth8x32} is 0.56 and 0.28 mJy (the rms for data without $\nu_n^{-1/3}$ scaling was slightly better). Corresponding 3-$\sigma$ upper limits on the stacked spectra are 1.7 mJy and 0.84 mJy for 47 Tuc and $\omega$ Cen, respectively. These limits were confirmed by injecting fake signals with a  velocity dispersion of 275 km s$^{-1}$ at the corresponding Psn$\alpha$ line frequency into the individual, pre-stacked data. The fake signal injection also served to verify the data reduction pipeline. Corresponding stacked spectra at the frequencies of H recombination lines detection were also made, with no detection.

\begin{table*}
    \centering
    \begin{tabular}{lcccccc}
    \hline
    Target      & Distance $D$ & Flux density & Recombination rate $R_n$ & LDM density $\rho_{0}$ & LDM mass & $J_0$\\
                &    (kpc) &  (mJy) & (s$^{-1}$)  &  (M$_{\odot}$ pc$^{-3}$)  & (M$_{\odot}$)  &  GeV$^2$ cm$^{-5}$  \\
    \hline

    47 Tuc &  4.4 & $<1.7$ & $<1.4\times10^{43}$ & $<54f_n^{-0.5}$ &  $<1.3\times 10^5f_n^{-0.5}$ & $<2.6\times 10^{21}f_n^{-1}$ \\
    
    $\omega$ Cen & 5.5 & $<0.84$ & $<1.1\times10^{43}$ & $<48f_n^{-0.5}$ & $<1.2\times 10^5f_n^{-0.5}$ & $<1.3\times 10^{21}f_n^{-1}$ \\
    \hline
   
    \end{tabular}
    \caption{3-$\sigma$ upper limits for 47 Tuc and $\omega$ Cen at the assumed distances for the flux density, Ps recombination rate, rms density, mass and $J$-factor due to self-annihilating light dark matter (LDM). A Doppler-broadened velocity dispersion of 275 km s$^{-1}$ is assumed. The flux density and rate limits apply to all Ps recombination line emission at the cluster recession velocity within the frequency range considered and within the (frequency-dependent) Parkes beam. The rms density and mass limits refer to LDM within the scale radius $r_s$, assumed to be 10 pc, for an NFW-like profile. The $J$-factor is the integral of $\rho^2$ along the line of site and across the solid angle subtended by the cluster out to the projected scale radius. A particle mass of 1 MeV~c$^{-2}$ and an annihilation cross section $\langle\sigma v\rangle = 3\times 10^{-29}$ cm$^3$ s$^{-1}$ are assumed. $f_n$ is the ratio of the recombination rate to the annihilation rate. Scaling for other values is given in Equation~\ref{eq:rmsdensity} for density and mass, and in Equation~\ref{eq:Jfactor} for the $J$-factor.
    }
    \label{limits}
\end{table*}

\section{DISCUSSION}
\label{sec:discussion}

The rms noise in the final spectra for 47 Tuc and $\omega$ Cen shown in Figure~\ref{fig:smooth8x32} is an order of magnitude higher than expected from the radiometer equation: 280-560 $\mu$Jy versus $40\mu$Jy expected for the given integration time, receiver noise temperature, number of polarisations, number of stacked lines, final spectral resolution (1 MHz), and observing technique (on-off). This seems mainly due to a combination of RFI, standing waves and receiver bandpass calibration. Despite the RFI, the spectral rms is $10^{10}$ times lower than the (continuously present) RFI maximum at 762 MHz (Figure~\ref{fig:47Tuc_TP_full}), an impressive spectral dynamic range.  
The 3-$\sigma$ flux density limit for $\omega$ Cen is also over two orders of magnitude better than the Galactic Centre measurement of \citet{Reynolds_2017}.

The heliocentric radial velocities for 47 Tuc and $\omega$ Cen are $-18$ km s$^{-1}$ and 232 km s$^{-1}$, respectively \citep{Harris1996}. Correcting for the telescope velocity at the time of observation, these correspond to topocentric radial velocities of $-5$ km s$^{-1}$ and 211 km s$^{-1}$, respectively. In turn, these correspond to the frequency offsets (for their respective reference frequencies of 2723.61 and 2648.31 MHz) of 0.05 MHz and $-1.86$ MHz, as shown by the dashed vertical lines in Figure~\ref{fig:smooth8x32}. For a gas temperature of $5\times 10^3$ K, the expected Ps thermal velocity dispersion is $\sim275$ km s$^{-1}$, corresponding to a FWHP linewidth of 5.8 MHz at 2.7 GHz, shown by the vertical orange bands in Figure~\ref{fig:smooth8x32}. 

Upper limits to the recombination line flux density and the corresponding recombination rates are listed in Table~\ref{limits}. The 3-$\sigma$ recombination rate upper limits vary from $R_n=1.1$ to $1.4\times10^{43}$ s$^{-1}$, with the closer distance of 47 Tuc slightly compensating for the somewhat poorer flux density limits. These limits are 210--270 times better than recent radio recombination rate limits obtained for the Galactic Centre \citep{Reynolds_2017}.

However, these limits are still poorer than the point source limits established from INTEGRAL/SPI all-sky maps in unconfused regions away from the Galactic Plane. For $\omega$ Cen, \citet{Knodlseder2005} quote  a 511-keV point-source upper limit of  $1.7\times 10^{-3}$ photons cm$^{-2}$ s$^{-1}$. At the distance quoted in Table~\ref{limits}, this corresponds to a Ps-only annihilation rate limit of $1.2\times 10^{42}$ s$^{-1}$ (counting only 2-photon emission from para-Ps), which is  a factor of 10 better than our recombination rate limit.

The radio and $\gamma$-ray limits can be translated into limits on the presence of self-annihilating dark matter. However, the limits depend sensitively on the density profile of any dark matter present. This in turn depends on: (a) the extent of baryonic feedback which can smooth the central density cusp \citep{Benitez2019}; and (b) the mass of the central black hole which can give rise to dark matter density spikes \citep{Fortes2020}. For a pure NFW dark matter profile \citep{Navarro1996}, the density profile is parameterised by:

\begin{equation}
\rho(r) = \rho_{0} \left( \frac{r}{r_s} \right) ^{-1} \left( 1+ \frac{r}{r_s} \right)^{-2} .
\end{equation}
The corresponding annihilation rate, volume-integrated to radius $r_m$, is:
\begin{equation}
    R_P(r_m) = \frac{4\pi}{3} \rho_{\circ}^2 m_{\Chi}^{-2} \langle\sigma v\rangle  r_s^3 \left(1-\frac{r_s^3}{(r_s+r_m)^3}\right),
\label{eq:rate}
\end{equation}
where $m_{\Chi}$ is the particle mass and $\langle\sigma v\rangle$ is the annihilation cross section. Below the scale radius $r_s$, the integrated annihilation scales as $r$, with equal signal contribution from all spherical shells. But this flattens off near the the scale radius, with the rate within the scale radius being given by $R_p(r_s) = (7\pi/6) \rho_{\circ}^2 m_{\Chi}^{-2} \langle\sigma v\rangle  r_s^3$. The density for self-annihilating dark matter within can be further parameterised as:
\begin{equation}
\begin{aligned}
\left( \frac{\rho_{0}}{{\rm M}_{\odot} ~ {\rm pc}^{-3}} \right) = &
46.4 \left( \frac{m_{\Chi}}{1~ {\rm MeV~c^{-2}}} \right) \left( \frac{\langle\sigma v\rangle}{3.10^{-29}~ {\rm cm}^3 ~{\rm s}^{-1}} \right)^{-0.5}  \\
 & \times \left( \frac{r_s}{10 ~ {\rm pc}} \right)^{-1.5} \left( \frac{R_p}{10^{43} ~ {\rm s^{-1}}} \right)^{0.5}.
\end{aligned}
\label{eq:rmsdensity}
\end{equation}

For a scale radius of $r_s=10$ pc, which approximately corresponds to the effective optical radii of the two clusters \citep{Gascoine1956}, and an annihilation cross section of $\langle\sigma v\rangle=3\times 10^{-29}$ cm$^3$ s$^{-1}$, the corresponding density upper limits shown in Table~\ref{limits} are 54 $f_n^{-0.5}$ ($m_\chi/$MeV~c$^{-2}$) M$_{\odot}$ pc$^{-3}$ (2.1$f_n^{-0.5}$ ($m_\chi/$MeV~c$^{-2}$) TeV cm$^{-3}$) and 48$f_n^{-0.5}$ ($m_\chi/$MeV~c$^{-2}$) M$_{\odot}$ pc$^{-3}$ (1.8$f_n^{-0.5}$ ($m_\chi/$MeV~c$^{-2}$) TeV cm$^{-3}$) for 47 Tuc and $\omega$ Cen, respectively.
The total self-annihilating LDM mass within the scale radius is $<1.3\times 10^5 f_n^{-0.5}$ ($m_\chi/$MeV~c$^{-2}$) M$_{\odot}$ and $<1.2\times 10^5 f_n^{-0.5}$ ($m_\chi/$MeV~c$^{-2}$) M$_{\odot}$ for 47 Tuc and $\omega$ Cen, respectively. 

\begin{figure}
    \centering
    \includegraphics[width=1.1\textwidth]{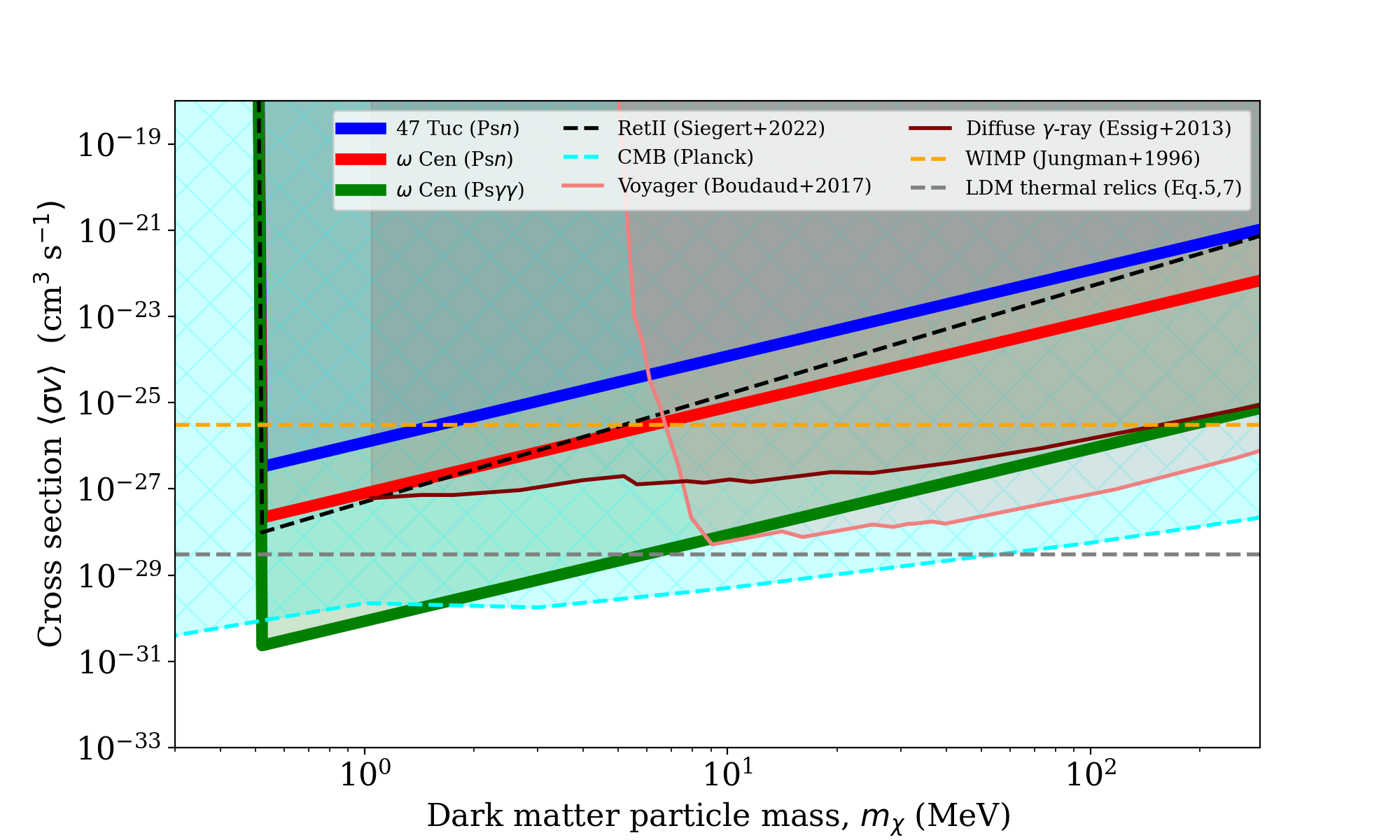}
     \caption{ Upper limits to the annihilation cross section $\langle\sigma v\rangle$, as a function of dark matter particle mass $m_{\chi}$. The solid blue and red lines (labelled Ps$n$) are the recombination upper limits for 47 Tuc and $\omega$ Cen, respectively, and assume a recombination/annihilation ratio $f_n\approx0.01$ and a central dark matter fraction of 5\%. The green line (labelled Ps$\gamma\gamma$) is the annihilation rate upper limit for $\omega$ Cen calculated from the INTEGRAL/SPI measurements of \citet{Knodlseder2005}, using the same dark matter fraction. The black dashed line is the Ps annihilation limit for Ret II \citep{SiegertRet2022}; the cyan dashed line is from Planck 2015/18 CMB anisotropy \citep{Slatyer2016,Kawasaki2021}; the solid pink line is from Voyager \citep{Boudaud2017}; the solid brown line is from the combined COMPTEL/INTEGRAL diffuse background \citep{Essig2013}. The horizontal orange and grey dashed lines are the canonical WIMP cross- ection \citep{Jungman1996} and the reference LDM thermal relic cross section (see text), respectively. } 
    \label{fig:xsection}
\end{figure}

The fiducial annihilation cross section of $\langle\sigma v\rangle=3\times 10^{-29}$ cm$^3$ s$^{-1}$ is chosen, following the arguments presented by \citet{Bartels2017} for LDM thermal relics, to be three orders of magnitude smaller than the canonical WIMP value cited by \citet{Jungman1996}. It
is also beneath the upper limit established for the nearby dwarf galaxy Reticulum II (Ret II) for all values of $m_{\chi}$ \citep{SiegertRet2022} -- see Figure~\ref{fig:xsection} for context. We will later discuss the implications for $\langle\sigma v\rangle$ if dynamical estimates of the dark matter content of the clusters are used instead.

For the smaller particle masses that we are most sensitive to (and the ones requiring least energy loss for Ps production),  $m_\chi \approx 1$MeV~c$^{-2}$, and assuming $f_n\approx 0.01$ for $n=106$, these limits correspond to 1.0 times the quoted stellar mass for 47 Tuc of $1.3\times 10^6$ M$_{\odot}$  \citep{Heyl2017} and 0.27 times the quoted stellar mass for $\omega$ Cen of  $4.5\times 10^6$ M$_{\odot}$ \citep{D'Souza2013}. 
For $\omega$ Cen, the $\gamma$-ray limits \citep{Knodlseder2005} give an even tighter upper limit to the LDM self-annihilating mass of 1\% of the stellar mass. 

The stripped dwarf galaxy simulations of \citet{Wirth2020} suggest that the expected density of dark matter could be an order of magnitude beneath our limit as a result of matter re-distribution near pericentric passage. Such interactions will evolve the dark matter profile shapes such that they are no longer NFW, even if they started out as NFW. However,  some of their models (e.g. S2) still retain substantial dark matter out to radii of $\sim 100$ pc, and therefore have dark matter masses higher than our limits in Table~\ref{limits}. 

The commonly-quoted astrophysical $J$-factor \citep{Charbonnier2011} is defined as:

\begin{equation}
    J = \int \int  \rho^2(x,\Omega) dx d\Omega , 
\end{equation}
where $x$ is the position along the line-of-sight and $\Omega$ is the two-dimensional angular coordinate.
For our fiducial NFW profile, the integral over all directions is $J = (4\pi/3) r_s^3 \rho_0^2/D^2$, where $D$ is the cluster distance. The majority of the double integral (92\%) is enclosed within the solid angle defined by the projected scale radius $r_s$. Hence:
\begin{equation}
\begin{aligned}
    \left( \frac{J_{0}}{{\rm M^2_{\odot}} ~ {\rm pc}^{-5}} \right) = & 0.33
    \left( \frac{m_{\Chi}}{1~ {\rm MeV~c^{-2}}} \right)^2 \left( \frac{\langle\sigma v\rangle}{3.10^{-29}~ {\rm cm}^3 ~{\rm s}^{-1}} \right)^{-1} \\
 & \times \left( \frac{D}{5 ~ {\rm kpc}} \right)^{-2} \left( \frac{R_p}{10^{43} ~ {\rm s^{-1}}} \right),
\end{aligned}
\label{eq:Jfactor}
\end{equation}
which, for $m_{\Chi}=1~ {\rm MeV~c}^{-2}$ and $\langle\sigma v\rangle = 3\times 10^{-29} {\rm cm}^3 {\rm s}^{-1}$, results in upper limits of $0.58 f_n^{-1}$ M$^2_{\odot}$ pc$^{-5}$ ($2.6\times10^{21}f_n^{-1}$ GeV$^2$ cm$^{-5}$) for 47 Tuc and $0.29 f_n^{-1}$ M$^2_{\odot}$ pc$^{-5}$ ($1.3\times10^{21} f_n^{-1}$ GeV$^2$ cm$^{-5}$) for $\omega$ Cen. Since $R_p \propto D^2$, the  distance-dependence in Equation~\ref{eq:Jfactor} disappears.

Our upper limits are based on an NFW profile with a scale radius $r_s=10$ pc. Other authors have considered more generic dark matter distributions \citep[e.g.][]{Evans2016} but, given the ongoing tidal interactions with the Galaxy, it is unknown which of these distributions might be more realistic. In particular, for a given mass, less cuspy models, which may be more consistent with observations of low-mass galaxies \citep{Oh2011}, result in fairly similar outcomes. Examination of Equation~\ref{eq:rate} shows that, even for a uniform sphere with density $\rho_0$, similar limits would result. More importantly, the FWHP beamwidth of the Parkes telescope \citep{Hobbs2020} ranges from $0.11^{\circ}$ at the rest frequency of Ps93$\alpha$ (4024.99 MHz) to $0.27^{\circ}$ at the rest frequency of Ps135$\alpha$ (1322.42 MHz), beyond which we have no sensitivity to Ps recombinations. However, this beamwidth, which corresponds to 10--23 pc at the average distance of 47 Tuc and $\omega$ Cen, is conveniently similar to, or slightly greater than, the effective radii of the clusters. If the dark matter is more concentrated, then the mass upper limits in Table~\ref{limits} are overestimated, and need to be reduced by a factor of $(r_s/10 \mathrm{pc})^{-3/2}$. But if more extended, our mass limits are underestimated and refer only to the mass contained within the Parkes beam. In principal, a very extended distribution ($>30$ pc for 47 Tuc; $>60$ pc for $\omega$ Cen) could result in signal cancellation in the reference (off-source) integration, but overall, the Parkes beam is well suited to the current experiment. It is much more useful for localisation than the INTEGRAL/SPI resolution of $2.5^{\circ}$ \citep{Winkler2003}, and has higher brightness sensitivity than current radio interferometers. 

LDM particles heavier than 1 MeV (up to 1 GeV) result in fewer annihilations, and poorer limits on dark matter density. On the other hand, Cosmic Microwave Background observational limits on the annihilation cross section $\langle\sigma v\rangle$ via the $\chi\chi\rightarrow e^+e^-$ channel  are also less severe by several orders of magnitude \citep[][see Figure~\ref{fig:xsection}]{Slatyer2016,Bartels2017,Dutra2018,Kawasaki2021}.

As discussed in Section~\ref{sec:positronium}, the limits presented in Table~\ref{limits} are also only relevant where $e^+e^-$ annihilation occurs via the Ps channel, which requires low energies and low temperatures, and therefore energy loss within the interstellar medium (ISM). Without such moderation, the favoured process is direct annihilation. For future work, this favours observations of nearby dark matter-rich galaxies which have an ISM but which are without an active nucleus or active star-forming regions. The class of dwarf galaxies around the Milky Way satisfy some of these criteria. In particular, the ultra-faint dwarf galaxy Ret II, discovered by the Dark Energy Survey team \citep{Bechtol2015}, stands out as a good candidate. Observations and numerical simulations demonstrate that Ret II must contain a large amount of dark matter \citep{Simon2015, Armstrong2021}.  Fermi-LAT has detected $\gamma$-ray emission \citep{Hooper_2015} which may originate from the annihilation of dark matter particles \citep{Regis_2017} since the low stellar density environment of Ret II makes it a less-ideal environment than globular clusters to host re-cycled millisecond pulsars. Although initial analysis of INTEGRAL/SPI data suggested the presence of a 511 keV annihilation line at a significance of 3.1$\sigma$ \citep{SiegertSat2016}, this has since been retracted \citep{SiegertRet2022}. The challenge with dwarf galaxies is their much lower $J$-factors. For example, \cite{SiegertRet2022} quote a range of $0.2-3.7\times10^{19}$ GeV$^2$ cm$^{-5}$ for Ret II, whereas \citet{Evans2022} derive a probable range $10^{22-24}$ GeV$^2$ cm$^{-5}$ for $\omega$ Cen, similar to the upper limit from Equation~\ref{eq:Jfactor} of $1.3\times 10^{23}$ GeV$^2$ cm$^{-5}$ for $f_n=0.01$, $m_{\Chi}=1~ {\rm MeV~c}^{-2}$ and $\langle\sigma v\rangle = 3\times 10^{-29} {\rm cm}^3 {\rm s}^{-1}$. 

A corollary of our upper limit to the self-annihilating LDM mass for the two clusters is that, if the LDM mass is to be dynamically significant (i.e.\ similar to the stellar mass), then the annihilation cross section $\langle\sigma v\rangle < 3.1\times 10^{-31} f_n^{-1} ($m$_\chi/1$ MeV~c$^{-2})^2$ cm$^3$ s$^{-1}$ for 47 Tuc and $< 2.0\times 10^{-32} f_n^{-1} ($m$_\chi/1$ MeV~c$^{-2})^2$ cm$^3$ s$^{-1}$ for $\omega$ Cen. More realistic estimates from observations and numerical simulations suggest lower dark matter masses, and therefore less-constrained cross sections. The GAIA, HST and spectroscopic data analysed by \citet{Evans2022} suggest a half-light dark mass of $<10^6$ M$_{\odot}$ for $\omega$ Cen. However, analysis of observations is also affected by the treatment of high-velocity stars \citep{Lane2010,Claydon2019}. The majority of the simulations by \citet{Wirth2020} suggest central dark matter fractions in the range 0.01 to 0.1. Adopting a dark matter fraction of 0.05 for both clusters (i.e.\ M$_{\rm dark}=6.5\times10^4$ M$_{\odot}$ and $2.2\times10^5$ M$_{\odot}$ for 47 Tuc and $\omega$ Cen, respectively), more realistic estimates for the annihilation cross section are $\langle\sigma v\rangle <1.2\times 10^{-28} f_n^{-1} ($m$_\chi/1$ MeV~c$^{-2})^2$ cm$^3$ s$^{-1}$ for 47 Tuc and $<7.9\times 10^{-30} f_n^{-1} ($m$_\chi/1$ MeV~c$^{-2})^2$ cm$^3$ s$^{-1}$ for $\omega$ Cen. 
For $\omega$ Cen, the existing $\gamma$-ray limits \citep{Knodlseder2005} give a cross section limit of $\langle\sigma v\rangle < 8.6\times 10^{-31} ($m$_\chi/1$ MeV~c$^{-2})^2$ cm$^3$ s$^{-1}$. These results are shown in Figure~\ref{fig:xsection}.

For $f_n\approx0.01$, the cross-section limit derived from Ps recombination for $\omega$ Cen is similar to the $\gamma$-ray limit for Ret II \citep{SiegertRet2022}, also shown in Figure~\ref{fig:xsection}. However, the cross-section limit derived from Ps annihilation in $\omega$ Cen is over 2 orders of magnitude tighter, by virtue of the vastly larger $J$-factor for globular clusters compared with even nearby dwarf galaxies. Between 0.5 and 2 MeV the limits improve on those obtained for the same annihilation channel from CMB anisotropy results \citep{Slatyer2016,Kawasaki2021}, as also shown in Figure~\ref{fig:xsection}.

Finally, as already noted, whilst the radio recombination limits presented in Table~\ref{limits} are useful, they are an order of magnitude poorer than theoretically possible with the observing time that was available, and possibly 2 orders of magnitude less sensitive (in terms of annihilation sensitivity) than the best $\gamma$-ray observations. The reasons for this are the limitations in our ability to perform spectral calibration in the presence of several effects: (a) differing power levels in the target and references fields; (b) a telescope standing wave with a fundamental of 5.7 MHz which is too close to the width of the lines we are trying to detect (see Section~\ref{sec:analysis}); (c) ubiquitous radio frequency interference; and (d) the sub-division of the spectrum into 128-MHz sub-bands which created a large-amplitude 64-MHz pseudo-periodicity. 

\citet{Reynolds_2017} have shown that phased-array feeds have the advantage of a wider field of view and almost complete elimination of the standing wave. It is therefore possible that, with a combination of the cryogenic phased array feed currently under construction for the Parkes telescope and more accurate observational and calibration techniques, sensitivity to broad lines can be improved by at least an order of magnitude. Larger telescopes such as FAST, or SKA precursor telescopes on radio-quiet sites, may provide alternative avenues for exploration. If achievable, radio recombination line observations may be a useful alternative to $\gamma$-ray spectroscopy when greater angular resolution is required. Greater sensitivity should also be possible using high-surface brightness sensitivity spectroscopic observations of lower-$n$ lines in the optical to THz regimes.

\section{SUMMARY AND FUTURE WORK}
\label{sec:summary}

We have obtained observations with the 64-m Parkes telescope (Murriyang) of the Galactic globular clusters 47 Tuc and $\omega$~Cen in order to measure the amplitude of any positronium (Ps) recombination-line radiation, predicted to arise in some Light Dark Matter (LDM) models where the particles self-annihilate. 

Upper limits to the recombination rates { of $\sim 10^{43}$ s$^{-1}$ were} obtained for both clusters. Using modest values for the likely annihilation cross section  ($\langle\sigma v\rangle=3\times 10^{-29}$ cm$^3$ s$^{-1}$), upper limits to the { rms dark matter density and mass of  $\sim 50 f_n^{-0.5}$ (m$_\chi/$MeV~c$^{-2}$) M$_{\odot}$ pc$^{-3}$ and $\sim 1.2\times10^5 f_n^{-0.5}$ (m$_\chi/$MeV~c$^{-2}$) M$_{\odot}$ were obtained, respectively}. For $\omega$~Cen, this corresponds to a limit on the self-annihilating light { dark matter mass of $<30$\% of the stellar mass, assuming a recombination/annihilation ratio, $f_n\approx 0.01$, and a particle mass of 1~MeV~c$^{-2}$. Archival INTEGRAL/SPI observations suggest an even smaller limit, $<1$\% of the stellar mass}.

Flipping the problem around and using available dynamical models and observational data which suggest a plausible dark-to-luminous ratio of 5\% within the half-light radius of the clusters, we can alternatively derive a dark matter annihilation cross section $\langle\sigma v\rangle <7.9\times 10^{-30} f_n^{-1} ($m$_\chi/$MeV~c$^{-2})^2$ cm$^3$ s$^{-1}$ for $\omega$ Cen from our new recombination results, and $\langle\sigma v\rangle < 8.6\times 10^{-31} ($m$_\chi/$MeV~c$^{-2})^2$ cm$^3$ s$^{-1}$ from an analysis of archival $\gamma$-ray limits \citep{Knodlseder2005}. The former limits are similar to the recent $\gamma$-ray limits derived for Reticulum II \citep{SiegertRet2022} and, for MeV-scale dark matter, the latter limits are stronger than those derived from CMB anisotropies \citep{Slatyer2016,Kawasaki2021}.

The { recombination} limits were possible by using the capability of a new ultra-wide-bandwidth low-frequency receiver which enabled sensitive, simultaneous observations over 0.704 -- 4.032 GHz, a frequency range which contains 73 Psn$\alpha$ recombination lines, though limitations caused by radio frequency interference and coarse filter artifacts only allowed us to stack only { a third of these} lines simultaneously.

Directions for further radio spectral-line observations include observations of nearby dark matter-dominated dwarf galaxies, particularly those around the Milky Way that have an interstellar medium for moderation of $e^+$ particles. { Reticulum II \citep{Bechtol2015} is one such candidate. It is preferable to investigate sources without sources of emission associated with strong star-formation or black hole activity. Nevertheless, the presence of black holes will produce dark matter density spikes that can substantially boost signal}. 

Thermal recombination lines are usually stronger at higher frequencies, so further investigation at millimeter, THz or even infrared wavebands 
may be fruitful, as long as high surface brightness sensitivity is available. Due to better spatial resolution, these may complement $\gamma$-ray searches. Although low predicted values for the recombination/annihilation ratio disfavours the chances of Ps line detection at radio wavelengths, it is always possible that maser activity due to inverted populations, or absorption against background radiation could boost recombination line strengths by orders of magnitude.

\begin{acknowledgement}
Thanks to the anonymous referee for their helpful comments, and thanks to Ella Wang, Tristan Reynolds, Jonghwan Rhee, Ron Ekers and Roland Crocker who contributed to an earlier precursor project.
The Parkes radio telescope is part of the Australia Telescope National Facility which is funded by the Australian Government for operation as a National Facility managed by CSIRO. We acknowledge the Wiradjuri people as the traditional owners of the Observatory site. Parts of this research were supported by the Australian Research Council Centre of Excellence for All Sky Astrophysics in 3 Dimensions (ASTRO 3D), through project number CE170100013.
\end{acknowledgement}

\bibliography{globulars.bib}

\end{document}